\documentclass{article}
\usepackage{spconf,amsmath,graphicx}
\usepackage{amsfonts}  
\usepackage{amssymb}                 
\usepackage{dsfont}

\usepackage{bm}
\usepackage[linesnumbered,ruled,vlined]{algorithm2e}  
\SetKwInput{KwInput}{Input}
\SetKwInput{KwProcess}{Process}
\SetKwInput{KwOutput}{Output}          
\usepackage{multirow}                

\usepackage{amssymb}                 
\usepackage{color}
\usepackage{multicol}                
\usepackage{setspace}
\usepackage{lipsum}
\usepackage{booktabs}

\title{A Low-Complexity ADMM-based Massive MIMO Detectors via Deep Neural Networks}
\name{Isayiyas Nigatu Tiba, Quan Zhang, Jing Jiang and Yongchao Wang}
\address{State Key Laboratory of Intergrated Services Networks\\
 Xidian University, Xi'an, China}
\begin{document}
%
\maketitle
\newcommand\blfootnote[1]{%
\begingroup
\renewcommand\thefootnote{}\footnote{#1}%
\addtocounter{footnote}{-1}%
\endgroup
}
\begin{abstract}
An alternate direction method of multipliers (ADMM)-based detectors can achieve good performance in both small and large-scale multiple-input multiple-output (MIMO) systems. However, due to the difficulty of choosing the optimal penalty parameters, their performance is limited. 
This paper presents a deep neural network (DNN)-based massive MIMO detection method which can overcome the above limitation. It exploits the unfolding technique and learns to estimate the penalty parameters. Additionally, a computationally cheaper detector is also proposed. The proposed methods can handle the higher-order modulation signals.  Numerical results are presented to demonstrate the performances of the proposed methods compared with the existing works.      
\end{abstract}
\vspace{-10pt}
\section{Introduction}\label{intro.}
Several deep neural networks (DNN)-based detectors are proposed in the literature to improve the efficiency of existing traditional multi-input multi-output (MIMO) detectors \cite{samuel2019detect} \cite{hemodel-driven}. These detectors are designed by using the ``{\it deep unfolding}" technique \cite{deepunfolding}. This technique considers the iterations of existing algorithms to be a network layer; and then learns from the data to estimate optimal parameters for the underlying algorithm. For example, the work in \cite{samuel2019detect} proposed a network called ``{\it DetNet}" that unfolds the iterations of the projected gradient descent. Similar idea is used in the works of  \cite{adaptive}--\hspace{-0.008cm}\cite{wei2020learned}.

Further, detectors based on the alternating direction multiplier (ADMM) algorithm are also proposed, both in the traditional and deep learning approaches. In such cases, since ADMM is used to solve the optimal maximum likelihood (ML) problem with some constraints, the resulting detectors can achieve a sub-optimal performance.  For instance, the authors of \cite{admm-infintynorm}  proposed an infinity norm constrained traditional detector known as ``{\it ADMIN}" based on the ADMM.  Further, the work in \cite{admm-net} proposed a network called ``{\it ADMM-Net}" by unfolding the iterations of the ADMM algorithm.   The ADMM-Net has a similar problem structure as the ADMIN, but it is limited to the BPSK and QPSK modulations. Furthermore, recently, another ADMM-based detector known as ``{\it PS-ADMM}" is proposed in \cite{zhang2020designing}. In that work, the authors have proposed a penalized ML detection formulation be solved via the sharing-ADMM method that can handle high-order QAM signals efficiently.
They have also shown that the theoretical performance of the PS-ADMM is guaranteed. However, since there is no analytical method for obtaining the optimal penalty parameters \cite{wohlberg2017admm} \cite{boyd2011distributed}, it is challenging to improve the performance in the approaches mentioned above. 

Inspired by the work of \cite{admm-net} and \cite{zhang2020designing}, and using the unfolding method, we propose an efficient DNN-based ADMM detector. Our work extends the existing algorithm of \cite{zhang2020designing} in two different forms. First, we unfold the iterations of the PS-ADMM and design a DNN architecture that can efficiently approximate the optimal penalty parameters. Second, we customize the standard multilayer perceptron (MLP) \cite{hornik1989multilayer} into the ADMM and utilize different feature selection strategies to obtain improved performance with lower computational complexity.
\section{System Model and Problem Formulation}
We consider a  MIMO system in which a base station (BS) is equipped with $M_c$ antenna arrays to serve a user terminal equipped with $K_c$ number of antennas, where $M_c >K_c$. We assume that a transmit symbol vector $\tilde{{\bf s}}$ is drawn  from a set $\mathcal{S}_c$ in a rectangular $4^q$-QAM complex alphabets, where $q$ is a positive integer.  Then the complex received signal vector $\tilde{{\bf y}}$  can be written as 
\begin{align}
\tilde{\mathbf{y}} = \tilde{\mathbf{H}}\tilde{\mathbf{s}} + \tilde{\boldsymbol{\upsilon}}, \label{compModel}
\end{align} 
where   ${\tilde{\boldsymbol{\upsilon}}} \in  \mathbf{\mathbb{C}}^{M_c}$ is a zero mean, unit variance complex additive white Gaussian noise (AWGN), and $\tilde{\mathbf{H}}$ is a flat fading i.i.d Rayleigh channel. We transform \eqref{compModel} to an equivalent real model 
\begin{align}
\mathbf{y} = \mathbf{H}\mathbf{s} + \boldsymbol{\upsilon}, \label{real-model}
\end{align} 
where 
\small\begin{align*}
\mathbf{s}\!=\!\begin{bmatrix}
{\rm Re}(\tilde{\mathbf{s}})\\ {\rm Im}(\tilde{\mathbf{s}})
\end{bmatrix}\
\mathbf{y}\!=\!\begin{bmatrix}
{\rm Re}(\tilde{\mathbf{y}})\\ {\rm Im}(\tilde{\mathbf{y}})
\end{bmatrix}\
\mathbf{H}\!=\!\begin{bmatrix}
{\rm Re}(\tilde{\mathbf{H}})\ &{\rm Im}(\tilde{\mathbf{H}}) \\
-{\rm Im}(\tilde{\mathbf{H}})\ &{\rm Re}(\tilde{\mathbf{H}})
\end{bmatrix},
\end{align*}
$\boldsymbol{\upsilon}\in\mathbb{R}^M$ is the AWGN. The real domain MIMO size and symbol sets can be written as   $(M,K) = (2M_c,2K_c)$, and $\mathcal{S} = \{{\rm Re}({\mathbf{\mathcal{S}}_c})\}\bigcup\{{\rm Im}({\mathbf{\mathcal{S}}_c})\}$ respectively.  


%
Generally for any given $4^q$-QAM signal in the set $\mathcal{S}$ the transmit symbol can be written as 
\begin{align}
\mathbf{s} = \sum_{i=1}^{q}2^{i-1}\mathbf{z}_i,\label{decompose}
\end{align} 
where ${\bf z}_i\in \{-1,1\}^K$ for all $i = 1,2,\cdots, q$. 
Now let us introduce an auxiliary variable ${\bf x}\in\mathbb{R}^K$, and consider  the binary vectors $\{{\bf z}_i\}$ as penalty terms. Then by relaxing the constraint of the classical ML problem, the
MIMO detection problem can be written as 
\begin{align}
\label{relaxation}
\begin{split}
&\underset{{\bf x},{\bf z}_i}{\min} \frac{1}{2}\| \mathbf{y}-\mathbf{H}\mathbf{x}\|^2_2 - \frac{1}{2}\sum_{i=1}^{q}\alpha_i\|\mathbf{z}_i\|^2_2\\
&\text{s.t.} \quad \mathbf{x} - \sum_{i=1}^{q}2^{i-1}\mathbf{z}_i = 0,\quad {\bf z}_i\in\mathcal{Q}^K
\end{split}
\end{align}
where  $\mathcal{Q} = [-1,1]  $ is the box constraint that relaxes the binary set $\{-1,1\}^K$, and $\alpha_i\geq0$ are the penalty parameters.  
The  augmented Lagrangian corresponding to \eqref{relaxation} can be written as
\begin{align}
\begin{split}
\mathcal{L}_\rho (\{{\bf z}_i|i=1,2,\cdots,q\}, \mathbf{x},\boldsymbol{\lambda}) = \frac{1}{2}\| \mathbf{y}-\mathbf{H}\mathbf{x}\|^2_2  - \frac{1}{2}\sum_{i=1}^{q}\alpha_i\|\mathbf{z}_i\|^2_2 \\
+\boldsymbol{\lambda}^T\left(\mathbf{x} - \sum_{i=1}^{q}2^{i-1}\mathbf{z}_i\right) + \frac{\rho}{2}\|\mathbf{x}-\sum_{i=1}^{q}2^{i-1}\mathbf{z}_i\|^2_2,
\end{split}
\label{LP}
\end{align}
where $\boldsymbol{\lambda}\in\mathbb{R}^K,\,\rho,\,T$ are the dual variable, penalty parameter, and transpose operation respectively. 
%
 By defining the scaled dual variable as ${\bf u}= \frac{\boldsymbol{\lambda}}{\rho}$, and solving \eqref{LP} through ADMM in its equivalent scaled-form, we can obtain the following sequence of solutions \cite{zhang2020designing}
\begin{subequations}
	\small	\begin{align}
	{\bf z}_i^{t+1} := &{\Pi}_{[-1,1]}\bigg(\frac{2^{i-1}\rho}{4^{i-1}\rho \nonumber-\alpha_i}\bigg({\bf x}^{t}-\sum_{j^\prime =1,i\neq 1}^{i-1}2^{j^\prime-1}{\bf z}_{j^\prime}^{t+1}\nonumber \\ &-\sum_{j=i+1,q>1}^{q}2^{j-1}{\bf z}_j^{t}+{\bf u}^{t}\bigg) \bigg), i = 1,2,\cdots,q\label{recon.}\\
	{\bf x}^{t+1} :=& \bigg({\bf H}^T{\bf H}+\rho{\bf I}\bigg)^{-1}\bigg({\bf H}^T{\bf y}\nonumber \\ &+\rho\bigg(\sum_{i=1}^{q}2^{i-1}\mathbf{z}^{t+1}_i -{\bf u}^{t }\bigg)\bigg) , \label{x-update}\\
	{\bf u}^{t+1} := &{\bf u}^{t}+{\bf x}^{t+1}-\sum_{i=1}^{q}2^{i-1}\mathbf{z}^{t+1}_i, 
	\end{align}
	\label{ps_admm.}
\end{subequations} 
where $t=1,2,\cdots,\tau$ is the number of iteration, $ {\Pi}(\cdot)$ is a projection.

From the above procedure, we can see that the detection framework in \eqref{ps_admm.} can achieve a better solution and theoretically-guaranteed convergence  \cite{boyd2011distributed}. However, one of the issues with this strategy is the choice of penalty parameters, as there is no theoretical procedure for choosing the optimal values \cite{wohlberg2017admm}. Besides that, since \eqref{x-update} involves the matrix inversion operation, it incurs higher computation cost, to the overall complexity of the detector. 

To overcome these challenges, we propose DNN-based solutions by transforming  \eqref{ps_admm.} into the deep network architecture.  Unlike the work in \cite{admm-net}, which is limited to BPSK and QPSK signals, our method can detect QAM signals in higher orders efficiently. In addition, we will present a low-complexity learning strategy for ADMM-based detection. 

\section{The proposed  DNN-based ADMM Detectors }
In this section, we present two distinct detectors based on the ADMM solutions \eqref{ps_admm.}. 

We begin by describing the deep unfolding procedure that can transform  \eqref{ps_admm.} into sequentially connected layers of DNN as follows.
Suppose that ${\bf w}^{\ell}_1\in\mathbb{R}^K$  denotes a linear transformation vector, where $\ell=1,2,\cdots,L$ is the number of layers. Then we define 
\begin{align}
	{\bf w}_i^{\ell+1} = \frac{2^{i-1}\rho}{4^{i-1}\rho \nonumber-\alpha_i}\bigg({\bf x}^{\ell}-\sum_{{j=1,i\neq1}}^{i-1}2^{j-1}{\bf z}_j^{\ell+1}\nonumber \\ -\sum_{j =i+1}^q2^{j-1}{\bf z}_{j}^{\ell}+{\bf u}^{\ell}\bigg), i = 1,2,\cdots,q\label{wi}
\end{align}
as linear transformations at the $\ell$-th layer, with a set of trainable parameters $\boldsymbol{\theta} = \{\{\alpha_i\}_{i=1}^q, \rho\}$, where $ {\bf x}^{\ell}, {\bf u}^{\ell}, \{{\bf z}_{j}^{\ell}\}^q_{j = i+1}$ are the inputs. 
Next, we untie  the projection $\Pi$ and define a pointwise activation function as
\begin{align}
	{\rm sgnlin}({\bf w}) = \min(\max(-1, {\bf w}), 1),\label{sgnlin}
\end{align}
where ${\bf w}\in\mathbb{R}^K$ is any vector. 

On the basis of the above transformations, we subsequently present the network architectures of the proposed detectors.  
\subsection{ADMM-PSNet} The goal of this network is to learn the set of optimal penalty parameters $\boldsymbol{\theta}$, from the data generated by the model \eqref{ps_admm.}. Its network architecture is constructed as follows.  
 \begin{subequations}
	\begin{align}
		\text{Initialize } {\bf x}^{0} =&{\bf 0}, {\bf u}^{0} = {\bf 0}, \{{\bf z}_{j}^{0}\}^q_{j = i+1} = {\bf 0}, i = 1,2,\cdots,q;\nonumber\\
		\text{For } \ell = &1,2,\cdots,L:\nonumber\\
		{\bf z}_i^{\ell+1} = &{\rm sgnlin}({\bf w}_i^{\ell+1}),\label{ps-zi} \\
		{\bf x}^{\ell+1}\big(\boldsymbol{\theta}\big) =& \bigg({\bf H}^T{\bf H}+\rho{\bf I}\bigg)^{-1}\bigg({\bf H}^T{\bf y}\nonumber\\ &+\rho\big(\sum_{i=1}^{q}2^{i-1}\mathbf{z}^{\ell+1}_i  -{\bf u}^{\ell }\big)\bigg),\label{ps-x}\\
		{\bf u}^{\ell+1} = &{\bf u}^{\ell}+{\bf x}^{\ell+1}\big(\boldsymbol{\theta}\big)-\sum_{i=1}^{q}2^{i-1}\mathbf{z}^{\ell+1}_i \label{ps-u}.
	\end{align}
\label{psnet-arc.}
\end{subequations}
This network is trained in an end-to-end fashion. In the forward pass, the parameter $\boldsymbol{\theta}$ is randomly initialized and kept constant in each layer. Then information propagates through each layer according to the update rules in \eqref{psnet-arc.}. At the output ( $L$-th layer), it computes  ${\bf x}^L$ and back propagates to update the parameter $\boldsymbol{\theta}$ by minimizing the loss function
\begin{align}
\mathcal{L}\big(\boldsymbol{\theta}\big) = \frac{1}{m}\sum_{i^\prime=1}^m\big\|{\hat{{\bf x}}}^{L(i)}\big(\boldsymbol{\theta}\big) -{\bf s}^{(i)}\big\|_2^2, \label{loss} 
\end{align}
through the stochastic gradient descent (SGD), where $m$ denotes the number of training examples. 
 This process will proceed until the acceptable error limit has been met. After the training is completed, ADMM-PSNet uses the trained parameters for online detection.

\subsection{ADMM-HNet}\label{admm-dnet}
This network introduces a hidden layer MLP in each iteration of \eqref{ps_admm.}. We begin the discussion by noting the computationally costly term $ \left({\bf H }^T{\bf H} + \rho{\bf I}\right)^{-1}$ in the ${\bf x}$-update \eqref{x-update}. Our goal is to obtain an accurate estimate of the ${\bf x}$-update without analytically computing this term by utilizing the approximation power of the MLP networks \cite{hornik1989multilayer}. To achieve this, first, we assume that the ${\bf x}$-update \eqref{x-update}  is unknown and describe it as a mapping function $\psi$: 
\begin{align}
{\bf x}^{\ell+1} = \psi^{\ell}\big({\bf y},{\bf H},  \{{\bf z}_{i}^{\ell+1}\}_{i=1}^q,\rho,{\bf u}^{\ell}\big),\quad \ell =1,2,\cdots, L .\label{x-mapping}
\end{align}
Then, to efficiently approximate the function $\psi^\ell$ we introduce an MLP network $g_n^\ell$  that has one hidden layer and $n$ number of units into the $\ell$-th layer. The key task here is to obtain a proper input (features) of the $g_n^\ell$ that leads to an effective learning environment without requiring costly computations. One way to do this would be to  combine all the inputs at each iteration; i.e. stacking the set ( ${\bf y},{\bf H}, {\rho}, \{{\bf z}_i^{\ell+1}\}_{i=1}^q,{\bf u}^\ell$ ) as a feature vector at each ${\bf x}$-update. However, since the received signal ${\bf y}$ and the channel ${\bf H}$ are constants through ADMM iterations, no new information will be provided by these parameters in each layer. Consequently,  the learning would be stuck, at some local minima, since ${\bf y}$ and ${\bf H}$ are dominating the input features. As such, in this work, we introduce the following procedure to obtain efficient training features.

Instead of directly using ${\bf y}$ and ${\bf H}$ in the input, let us introduce a residual variable ${\bf r}$,  defined as
\begin{align}
{\bf r}^{\ell+1} = {\bf y} - {\bf H}{{\bf x}}^{\ell},\label{residual}
\end{align} 
in each iteration.
Then we define a new feature vector ${\bf a}_1^{\ell}\in\mathbb{R}^K$ as
\begin{align}
{\bf a}_1^{\ell} = \frac{1}{M}\big({\bf H}^T{\bf r}^{\ell+1}\big) +{{\bf x}}^{\ell}. \label{a1}
\end{align}
Notice that the first term in \eqref{a1} works as successive noise cancellation, and the second term is the signal of interest from the previous layer. As the number of layers increases, since  ${\bf a}_1^{\ell}$ becomes more associated with the estimated signal,  it can be regarded as a subset of the potential variables to the input. In addition, we choose 
\begin{align}
{\bf a}^{\ell}_2 = \rho\big(\sum_{i=1}^{q}2^{i-1}\mathbf{z}^{\ell+1}_i  -{\bf u}^{\ell }\big),
\end{align} as another potential input variable since it naturally updates through the ADMM iterations. 
 Hence, the potential input feature set for the $\ell$-th layer can be obtained by vertically stacking the above features as
\begin{align}
{\bf a}^{\ell} = [{\bf a}_1^{\ell T}, {\bf a}_2^{\ell T}]^T. \label{input}
\end{align}
Notice that the dimension of each subset feature vector in \eqref{input} is $K\times 1$. Hence, the size of the input feature vector at each layer can be defined as $d_0 = 2K $. 

Given the above feature vector, the architecture of the $g_n^\ell$ can be described as follows.
Suppose that ${\bf W}_1^{\ell+1}\in\mathbb{R}^{n\times d_0},{\bf W}_2^{\ell+1}\in\mathbb{R}^{K\times n}$ and ${\bf b}^\ell_1\in\mathbb{R}^n, {\bf b}^\ell_2\in\mathbb{R}^K$ denote the weight matrices and the corresponding bias terms respectively. Then the forward pass of the $g_n^\ell$ at the $\ell$-th layer can be explicitly defined as   
\begin{align}
\begin{split}
 {\bf t}^{\ell+1} = \Gamma\big({\bf W}_1^{\ell+1}{\bf a}^\ell +{\bf b}_1^{\ell+1}\big),\,
\hat{{\bf x}}^{\ell+1} = {\bf W}_2^{\ell+1}{\bf t}^{\ell+1}+{\bf b}_2^{\ell+1},
\end{split}
\label{forwardpass}
\end{align}
where $\Gamma(.)$ is a pointwise ReLU activation: $\max(.,0)$, and ${\bf t}^{\ell+1}\in\mathbb{R}^n$ is an intermediate variable. Each network $g_n^\ell$ will be trained by minimizing the loss
\begin{align}
\mathcal{L}\big(\boldsymbol{\Theta}^{\ell+1}\big) = \frac{1}{m}\sum_{i^\prime=1}^m\big\|g_n^{\ell}({\bf a}^{\ell (i)};\boldsymbol{\Theta}^{\ell+1}) -{\bf s}^{(i)}\big\|_2^2, \label{loss2} 
\end{align} 
via SGD, where $\boldsymbol{\Theta}^{\ell+1} = \{ {\bf W}_1^{\ell+1}, {\bf W}_2^{\ell+1},{\bf b}_1^{\ell+1},{\bf b}_2^{\ell+1}\}$. Algorithm \ref{alg2} summarizes the training procedure of the ADMM-HNet.
\small\begin{algorithm}[h]
	\caption{\small Pseudo code fo the ADMM-HNet.}
	\label{alg2}
	\SetAlgoLined
	\KwInput{\small  ${\bf y}, {\bf H},L,q,m$,   $\{\{\alpha_{i}\}_{i=1}^q,\rho\}$; }
	Initialization: $\hat{{\bf x}}^{0} ={\bf H}^T{\bf y}/M, {\bf u}^{0} = {\bf 0}, \{{\bf z}_{j}^{0}\}^q_{j = i+1} = {\bf 0}$, $\boldsymbol{\Theta}^{\ell} =\{ {\bf W}_1^{\ell},{\bf W}_2^{\ell},{\bf b}_1^{\ell},{\bf b}_2^{\ell} \}^L_{\ell =1}$\;	
	\For{$\ell = 1,2,\cdots,L$ }{ \small
		Update  $\{{\bf z}_i^{\ell+1}\}$ via \eqref{ps-zi};\\  
		Generate the input features through \eqref{a1}-\eqref{input};\\	
		Construct the network $g_n^\ell$ according to \eqref{forwardpass};\\
		\For{$i^\prime = 1,2,\cdots, m$}{
			Construct the dataset ${\bf D} = \{{\bf a}^{\ell (i^\prime)}, {\bf s}^{(i^\prime)}\}$;\\
			Train the $g_n^\ell$ for some batchsize;
		}	
		Return $\hat{{\bf x}}^{\ell+1} = g_n^\ell\big({\bf a}^\ell, {\boldsymbol{\Theta}^{\ell+1}}\big)$;\\	
		Update the dual variable: ${\bf u}^{\ell+1} = {\bf u}^{\ell}+{\hat{{\bf x}}}^{\ell+1}-{\bf z}^{\ell+1}$;\\
	}
	\KwOutput{\small Return  $\hat{{\bf x}}^L$. }
\end{algorithm}

We conclude this discussion with the following technical remarks on penalty parameters and computational complexity. (1) In the ADMM-HNet above, since the penalty parameters do not appear in the network layers (see equation \eqref{forwardpass}), it is not convenient to learn them explicitly in the hidden layers. As such, we have pre-trained them in our implementation using the ADMM-PSNet model.
(2) The computational complexity of the ADMM-PSNet shall be the same as that of the PS-ADMM and ADMIN detectors. However, it is significantly reduced for the ADMM-HNet. From the procedures in section \ref{admm-dnet} and Algorithm \ref{alg2}, we can see that the complexity of this detector grows approximately in the order of $O(MK + L(MK +3Kn))$ where  $n$ is the number of units.
\section{Numerical Results}
In this section, different numerical results are presented to demonstrate the effectiveness of the proposed methods.
All detectors are implemented in python 3.6 and Keras API on top of TensorFlow. The ADMM-PSNet and ADMM-HNet are trained on $10,000 $ and $90,000$ for samples, respectively. 
The ADMM-PSNet is trained in an end-to-end manner with $L\!=\!30, { \rm epochs\!=\! 10,000}$, while the  ADMM-HNet, is trained in a layer-wise with $\rm batchsize\! =\! 1024,L\! =\! 30,\rm epochs\! =\! 2000$.
       
Fig.~\ref{q16layer} shows the generalization ability of our newly proposed ADMM-HNet in terms of its symbol-error-rate (SER) performance. 
{\setlength\abovedisplayskip{1.2pt}
	\setlength\belowdisplayskip{1.2pt}
	\begin{figure}[h]
		\centering
		\centerline{\includegraphics[width=6.80cm,height=4.8cm]{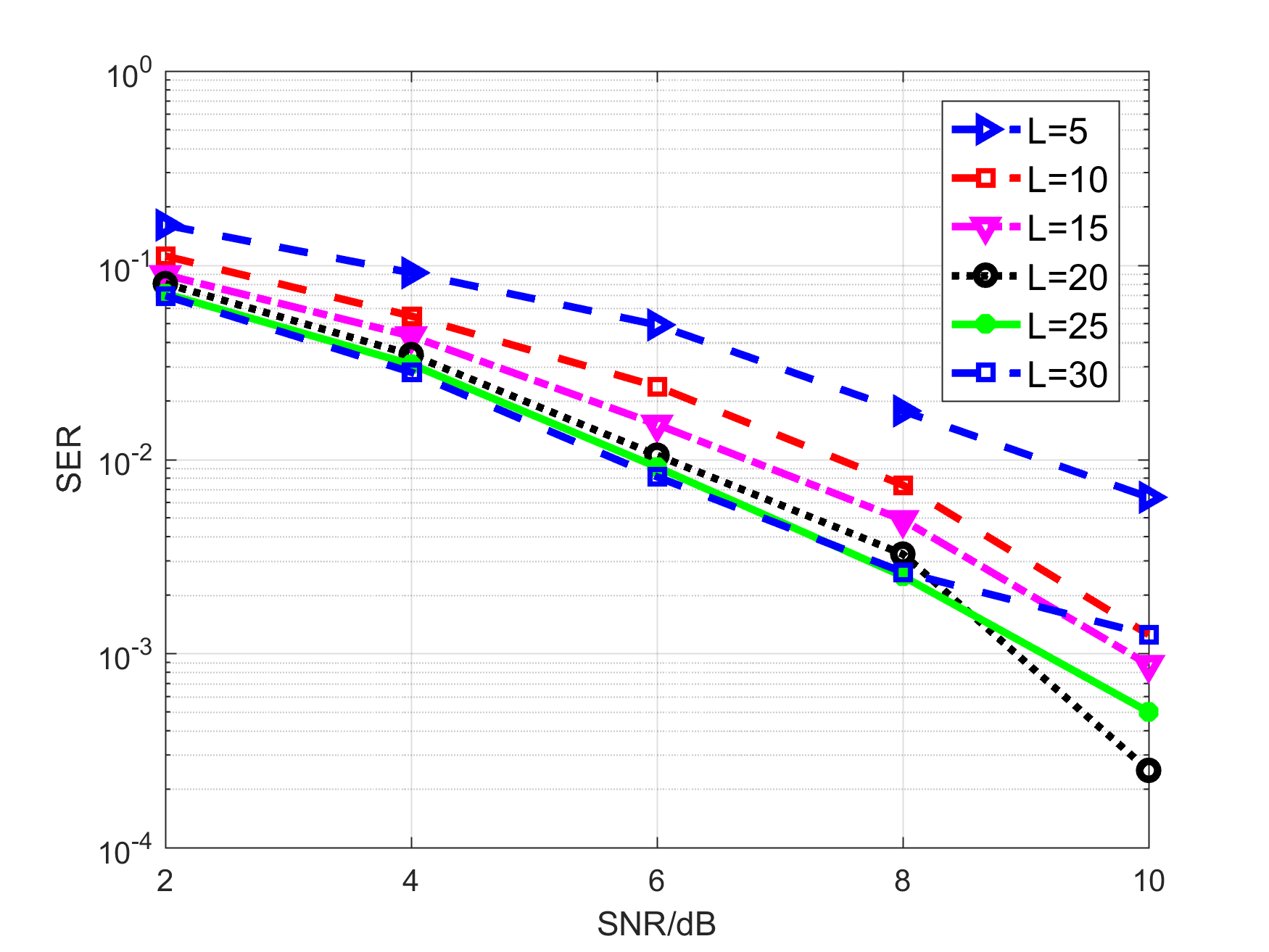}}
		\caption{\small The SER performances of the ADMM-HNet for a 16-QAM, $M = 64, K = 16$ for different number of layers.}
		\label{q16layer}
\end{figure}}
We run a Monte Carlo simulation on it  with a newly generated data to examine its performance at different layers. It is clear from the figure that as the number of  layers increases the performance also increased. Particularly when $L$ is in the range $1$ to $20$, the performance increased  significantly. On the other hand, when  $L$ increases beyond $20$, we can observe that it achieved nearly equal performances. This implies that, unlike some existing detectors, the ADMM-HNet does not require higher number of layers to achieve a competitive performance; i.e. with a properly selected input feature, and learned penalty parameters it can converge at early stage (such as $L = 20$).  
Next, to examine the effectiveness of the proposed methods, we  compare their performances with the existing detectors. 

Fig.~\ref{q16} illustrates the SER versus SNR performance comparison between the proposed and existing detectors for the 16-QAM, $M\!=\!64, K\! =\! 16$ MIMO system. As we can see the ADMM-PSNet performance is nearly similar to that of the PS-ADMM detector, whereas the ADMM-HNet outperformed the remaining detectors. Notice that the number of layers for this detector is 20, while the others run up to 30.
{\setlength\abovedisplayskip{1.2pt}
	\setlength\belowdisplayskip{1.2pt}
	\begin{figure}[h]
		\centering
		\centerline{\includegraphics[width=6.9cm,height=4.9cm]{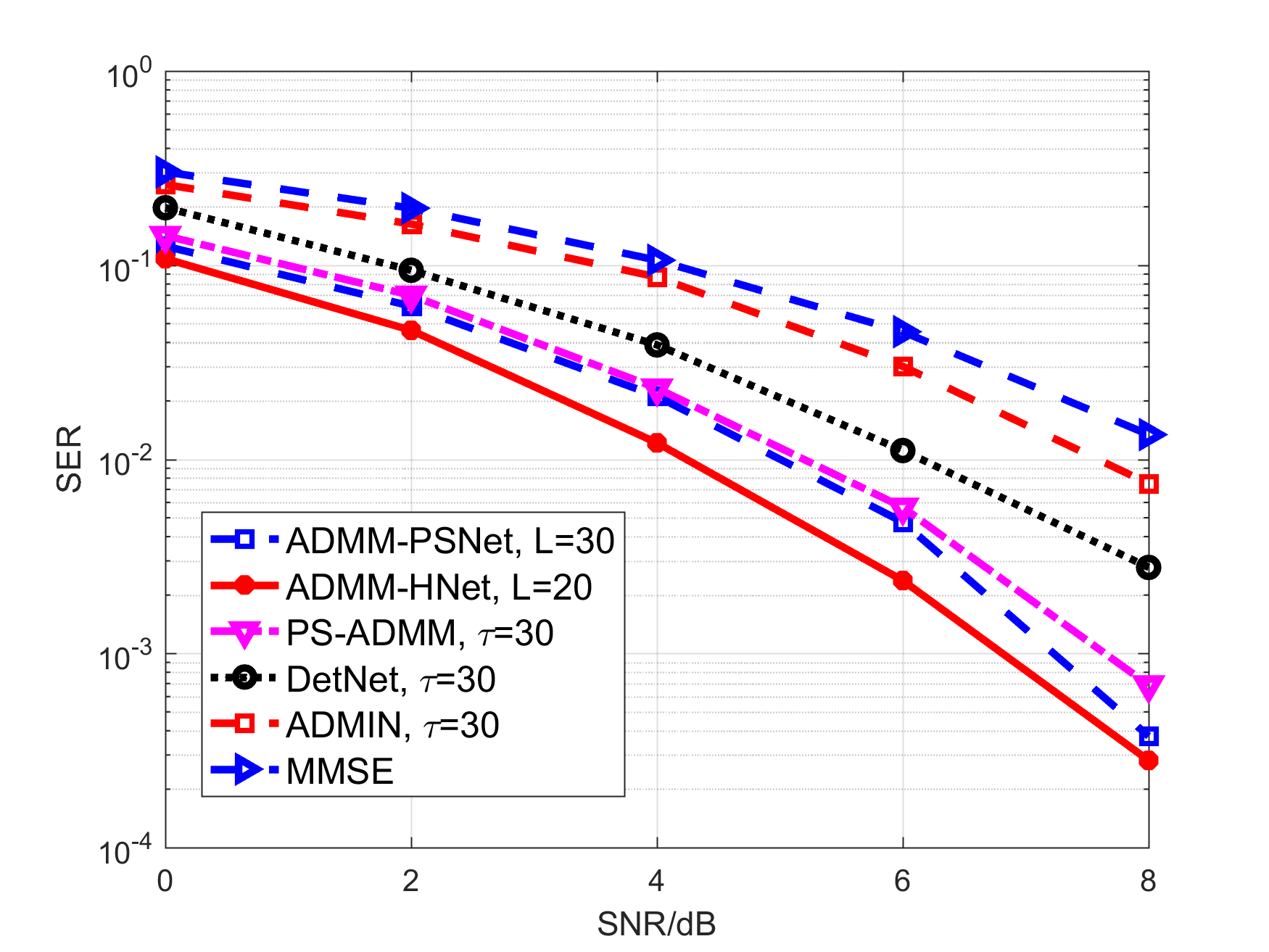}}
		\caption{\small The SER performance comparison for a 16-QAM, $M = 64, K = 16$.}
		\label{q16}
\end{figure}}
 
Fig.~\ref{q64} depicts the SER versus SNR performances of the detectors for the 64-QAM, $M\!=\! 64, K\!=\!16$ MIMO case. In this figure, we can see that the ADMM-PSNet achieved slightly better performance than the PS-ADMM under the higher SNR regime. On the other hand, ADMM-HNet performed almost similar to ADMM-PSNet and PS-ADMM under the lower SNR regime and improved performance under the higher SNR regime.
{\setlength\abovedisplayskip{1.2pt}
	\setlength\belowdisplayskip{1.2pt}
	\begin{figure}[h]
		\centering
		\centerline{\includegraphics[width=6.9cm,height=4.9cm]{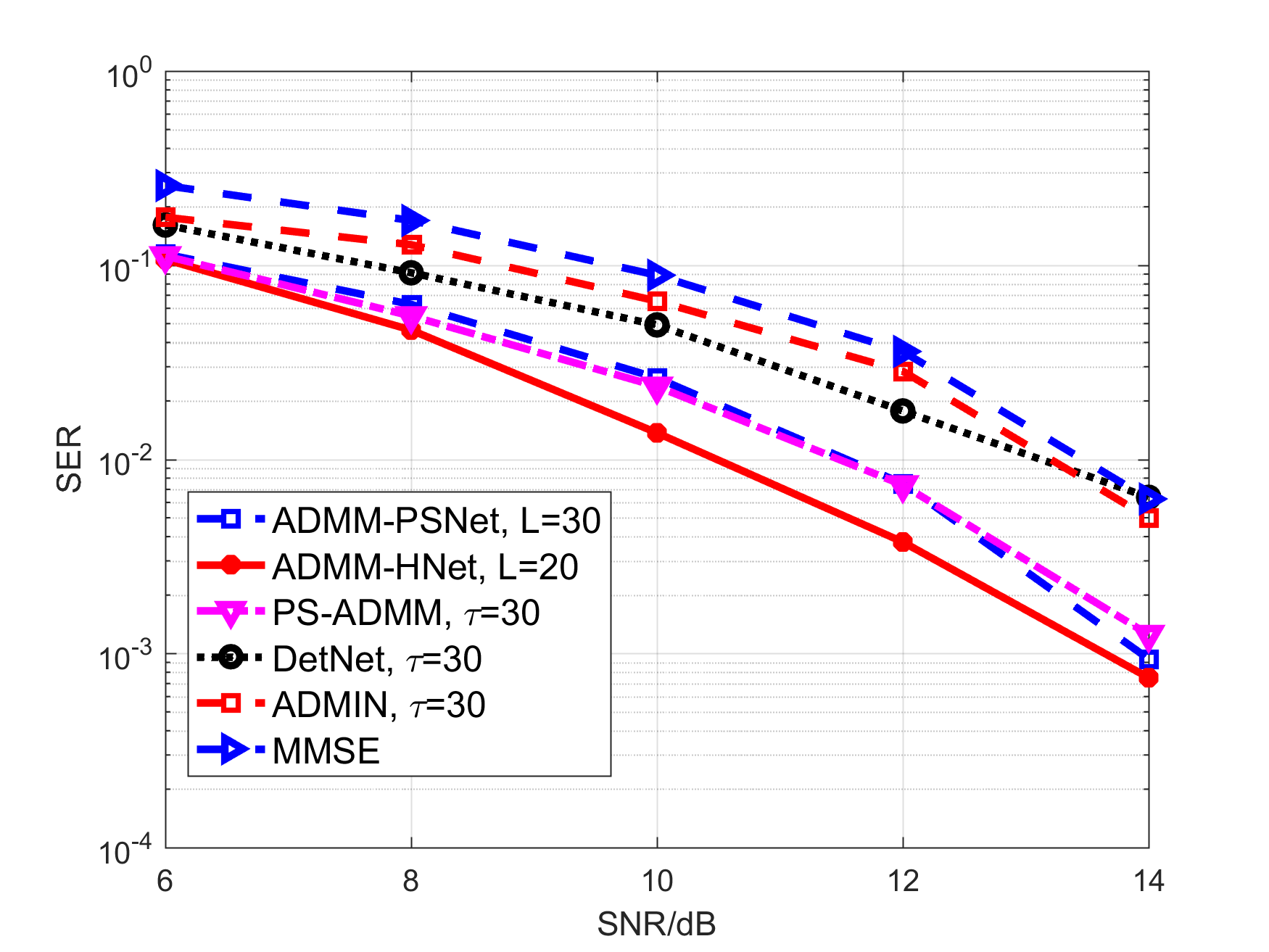}}
	\caption{\small The SER performance comparison  for a 64-QAM, $M\! =\! 64, K\! =\! 16$.}
	\label{q64}
\end{figure}}
These imply that with properly selected input features, the proposed ADMM-HNet can achieve a very competitive performance with a moderate number of layers such as 20. Moreover, as discussed in section \ref{admm-dnet}, since its computational complexity is cheaper compared to the others, this detector can efficiently serve the massive MIMO system. 

We conclude our discussion by comparing the average runtime between ADMM-based detectors in the table \ref{run-time}.\ These results are measured on the CPU: {\it Intel(R) Core(TM) i3-7100U CPU $3.90GHz\times 2$}.
\small\begin{table}[h]\centering
	\caption{\small  The average run time comparison measured in seconds;  ``*" stands for ``ADMM".}
	\label{run-time}
	\begin{tabular}{@{}lcccc@{}}\toprule
		Mod, $M\times K$	& \multicolumn{4}{c}{Detectors}\\
		\cmidrule{2-5} 
		& {\small *-HNet} & {\small PS-*} & {\small DetNet}&{\small ADMIN} \\ \midrule\midrule 
		{\small 16-QAM, $128\times 64$} &0.0029&0.0093&0.015&0.091  \\ 
		{\small 64-QAM, $64\times 16$} & 0.0018&0.0071&0.0083&0.076 \\ 
		\bottomrule 
	\end{tabular}
\end{table}
In this table, we can see that the ADMM-HNet reported faster time in both MIMO configurations. This is due to the fact that it is constructed from a cascade of a one-hidden-layer network that only computes a matrix-vector multiplication. The others, however, compute the multiplication and inversion of the matrix in each layer (iteration), and hence, they are relatively slower. 
\section{Conclusion}
In this paper,  DNN massive MIMO detection methods that can work with the higher-order QAM signals are proposed based on the ADMM algorithm. The ADMM-PSNet is trained to estimate the penalty parameters that result in improved performance. 
On the other hand, the ADMM-HNet is designed to reduce computational complexity by eliminating the matrix inversion and multiplication operations.  The numerical results reveal that the proposed detectors can significantly improve the performance of the existing methods. In Particular, the ADMM-HNet has shown improved performance with less computational complexity.


%

\end{document}